\documentclass[aps,prl,twocolumn,superscriptaddress,amssymb,showpacs
]{revtex4}
 \usepackage{amsmath}
\usepackage{graphicx}
\usepackage{epsfig}
\newcommand{\beq}{\begin{equation}}
\newcommand{\eeq}{\end{equation}}
\newcommand{\bea}{\begin{eqnarray}}
\newcommand{\eea}{\end{eqnarray}}
\newcommand{\pdag}{{\phantom{\dagger}}}

\begin{document}
\bibliographystyle{apsrev}
 
\title{Proposal of an experimentally accessible measure of many-fermion entanglement }
\author{M.\ Kindermann  }
\affiliation{ Laboratory of Atomic and Solid State Physics, Cornell University, Ithaca, New York 14853-2501  }

\date{November 2005}
\begin{abstract}
We propose an experimentally accessible measure of entanglement in many-fermion systems that characterizes  interaction-induced ground state correlations. It is formulated in terms of  cross-correlations of currents through resonant fermion levels weakly coupled to the probed system. The proposed entanglement measure vanishes in the absence of many-body interactions at zero temperature and it is related to measures of occupation number entanglement. We evaluate it for two examples of  interacting electronic nanostructures. 
  \end{abstract}
\maketitle

%Entanglement is an intrinsically quantum phenomenon and results  in correlations that defy any classical intuition. Its nonclassical nature is highlighted for instance by the Einstein-Podolsky-Rosen paradoxon and Bell's inequalities \cite{Bel64}. Entanglement is regarded 
Entanglement  is a distinguishing feature of quantum mechanics  \cite{Bel64} and it is regarded as  the defining resource for many of its modern applications such as quantum communication and quantum computation. Despite intensive efforts, however, a  thorough understanding of entanglement  beyond the   bipartite setting has not yet been reached. Its characterization   in systems of indistinguishable particles  has   proven particularly challenging. A key insight into the problem has been that entanglement is an observer dependent concept  \cite{Zan01,Bar03,Bar04}. Different characterizations of it are therefore possible and they may prove useful in different situations. 
A number of entanglement measures for systems of  indistinguishable fermions have thus been proposed  \cite{Zan02,Eck02,Bar03,Bar04,Shi03,Shi04} with focus on different aspects of the phenomenon. %Most of them are formulated in terms of occupation numbers of single particle states.
Studies of the entanglement of many-particle systems
%Applications of measures of entanglement of this kind to many-particle systems 
have already resulted in several valuable new insights \cite{Ost02,Vid03,Vid04} and they promise to continue to contribute to our understanding of  complex phenomena such as quantum phase transitions \cite{Ost02,Vid03}.   

Both, the challenge as well as the fascination of many-particle systems originate from many-body interactions.
Measures of the entanglement that is induced by these interactions \cite{Shi04} should therefore characterize such systems in a particularly instructive way. Their theoretical evaluation for systems of interacting fermions faces, however, severe limitations. 
   % Although also noninteracting fermionic systems can support entanglement \cite{Bee03}, the observer dependence of the concept of entanglement allows to construct measures that quantify the interaction-induced  entanglement only. Such measures that accordingly consider Slater determinants,  the eigenstates of noninteracting many-fermion Hamiltonians,    unentangled have been proposed in Refs.\  \cite{Zan02,Eck02,Bar03,Bar04,Shi03}.  
In space dimensions larger than one many of these  systems are neither  accessible by any analytical nor by any numerical \cite{Tro05} tools  available to date. Experiments, either directly on the systems of interest or on  quantum simulators, that is artificial systems with equivalent Hamiltonians, can  help to remedy the situation.   In this Letter we therefore propose an entanglement measure for many-fermion systems that can be evaluated with existing experimental techniques and that quantifies interaction-induced ground state correlations.   It is based on a generalization of the notion of entanglement put forward in Refs.\ \cite{Bar03,Bar04}. We show that it is closely related to measures of occupation number entanglement \cite{Zan02,Shi03} in case one can define a noninteracting counterpart of an interacting system, but that it generalizes the concept. Being experimentally accessible the proposed entanglement measure is a novel probe of  interaction-induced ground state correlations in fermionic systems.
 
To exemplify the proposed notion of entanglement we apply it to electronic nanostructures. Due to their small dimensions nanoscale conductors are typically strongly affected by many-body interactions and they are thus natural candidates for such studies. In fact, the characterization of interaction-induced correlations is one of the main challenges that these systems pose. Suitable experimental probes have been identified for some of these correlations, for instance  shot noise to detect fractional effective charges in quantum Hall systems \cite{Hei97}. For many nanostructures of interest, however, standard experimental techniques such as current correlation measurements are insensitive to interaction-induced correlations  \cite{Lev04}. The proposed many-fermion entanglement, in contrast, promises to be a systematic probe of such correlations in these systems. %Its measurement   selectively detects and characterizes quantum correlations that are due to many-body interactions. 
It involves the measurement of cross-correlations of electrical currents through resonant levels that are weakly coupled to the studied system. Both required ingredients,  tunable resonant levels in the form of quantum dots  as well as the ability to perform cross-correlation measurements of electrical currents \cite{Hen99}, have been demonstrated experimentally.
  
Formalizing the observer dependence  of entanglement Barnum, Knill, Ortiz, and  Viola have introduced the concept of a ``generalized entanglement''  \cite{Bar03}.  It is defined  in terms of a set of experimentally accessible observables $\mathfrak{h}=\{A_1\dots A_n\}$  rather than  a spatial partitioning of a  system. It reduces to the conventional measures of bipartite entanglement as one chooses for $\mathfrak{h}$  the set of all operators $A_j$ that act locally in  two partitions of a system. The degree of entanglement of a quantum state in this formulation is determined by the expectation values of the observables $A_j$ in that state. %One considers the convex cone $C$ of functionals spanned by the $\lambda_\alpha$ \cite{Bar03,Bar04}. 
A state is defined to  be entangled if it does not produce extremal expectation values.  
 This can be motivated by  the bipartite case where entanglement induces mixed reduced density matrices for the subsystems and thus renders  expectation values of  local operators generically non-extremal. 
%A simple example is the Bell-pair state $|\uparrow\rangle|\downarrow\rangle -|\downarrow\rangle |\uparrow\rangle$. The local spin expectation values $\mathfrak{1} \times \sigma_i$ vanish in this state, while they take their extremal expectation values $\pm 1/2 $ in unentangled states. 
%In order to focus on a finite number of experimentally accessible quantum states 
  More formally, a quantum state with density matrix $\rho_\alpha$  is  represented by a  linear functional $\lambda_{\alpha}$ on $\mathfrak{h}$ that gives the expectation values $\lambda_{\alpha}(A_j)={\rm tr}\,\rho_{\alpha} A_j$.  We employ a formulation of generalized entanglement based on convex cones \cite{Bar03} and consider the convex cone $C_n$ of all linear combinations $\lambda=\sum_{\alpha} p_{\alpha} \lambda_{{\alpha}}$, $p_{\alpha}\geq 0$ of  functionals  $\lambda_\alpha\in \{\lambda_{g},\lambda_0, \dots,\lambda_{n}\}$. The $\lambda_\alpha$ are defined by density matrices $\rho_\alpha$ of  the ground state ($\rho_g$) and  other experimentally accessible states ($\rho_j$) of a quantum system. Elements of $C_n$ with $\sum_\alpha p_\alpha =1$ are referred to as states. States that cannot be expressed as linear combinations of other states are extremal  in $C_n$ and imply extremal expectation values. They are thus called pure. 
 For  a state $\lambda\in C_n$ the degree of generalized entanglement ${\cal E}_n$  is defined through a Schur-concave function $S$ as \cite{footnote1}
 \beq \label{ent}
 {\cal E}_n(\lambda)= {\rm inf}\{S({\bf p})|\lambda=\sum_{\alpha}p_{\alpha} \lambda_{\alpha}\;\;{\rm with} \;\lambda_{\alpha}\;{\rm  pure}\;{\rm  states}\}.
 \eeq
 We take $S({\bf p})=-\sum_{\alpha} p_{\alpha} \ln p_{\alpha}$, the Shannon entropy.   ${\cal E}_n$  depends on both, the set of accessible observables $\mathfrak{h}$ as well as the set of states $\lambda_{\alpha}$ that define $C_n$. 
 
\begin{figure} 
\includegraphics[width=6.5cm]{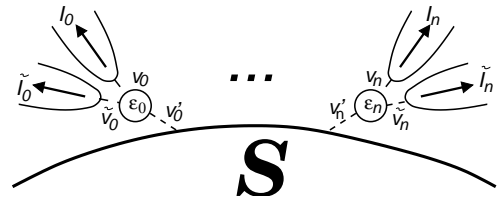}
%\vspace{4cm}
\caption{  System ${\cal S}$ contacted by resonant levels $j=0 \dots n$ that emit electron currents $I_j$, $\tilde{I}_j$ into two reservoirs each.}   \label{fig1}
\end{figure}

 Being formulated directly in terms of expectation values this generalized concept of entanglement  allows us to define the advertised experimentally accessible measure of  many-fermion entanglement. We introduce it for electronic systems ${\cal S}$, but it has an obvious extension to arbitrary fermionic systems. We choose electrical currents as the observables $A_j$. Typically electrical currents are correlated even in the absence of electron-electron interactions \cite{Bla00}.  To selectively characterize  the quantum correlations induced by many-body interactions we therefore consider the setup depicted in Fig.\ \ref{fig1}:  ${\cal S}$ is weakly coupled to at least two resonant levels $j$ with  resonance energies $\epsilon_j$. For small level broadenings $\Gamma_k,\Gamma_l \ll |\epsilon_k-\epsilon_l| $ the currents through two levels $j$, $k$ are then uncorrelated  in the absence of interactions.  $\Gamma_j$ are due to a coupling of the levels to two macroscopic leads (``reservoirs'') each, corresponding to the Hamiltonian
 \beq
 H= H_{0} + \sum_j  \epsilon_j d^\dag_j d^\pdag_j+\left[v'_j \psi_j^\dag d_j^\pdag +  d^\dag_j (v_j a^\pdag_j + \tilde{v}_j \tilde{a}^\pdag_j) + h.c.\right].
 \eeq
 Here,  $H_0$ is the Hamiltonian of ${\cal S}$ and the reservoirs when decoupled, $\psi_j^\dag$ creates an electron in ${\cal S}$ in the mode that is contacted by level $j$, $d^\dag_j$ creates an electron on level $j$, and $a^\dag_j$, $\tilde{a}^\dag_j$ create electrons in the reservoirs attached to it.  ${\cal S}$ is assumed to be  initially uncoupled from the reservoirs and in its ground state. After the coupling has been turned on it is still  arbitrarily close to its ground state  in the limit $v_j'\to0$ that we consider. We thus call the state of the coupled setup  $|g\rangle$.
%We  have $\Gamma_j=2\pi \nu(|v_j|^2+|\tilde{v}_j|^2)$, where $\nu$ is the density of states in the reservoirs. 
We refer to the currents into  the reservoirs coupled to $d_j$ as $I_j$ and  $\tilde{I}_j$ respectively.  We assume idealized noninteracting reservoirs with infinite bandwidth that are either completely occupied by electrons or entirely empty, such that $I_j\propto \langle g| f^\dag_j f^\pdag_j|g\rangle$. We have $f_j=d_j$ if the reservoirs coupled to $j$ are empty and $f^\pdag_j=d^\dag_j$ if they are occupied. Due to the Pauli principle $I_j$ and $\tilde{I}_j$ are perfectly anticorrelated at equal times: $\langle \tilde{I}_j(t) I_j(t)\rangle_{\rm s}=0$ while $\langle I_j\rangle,\langle \tilde{I}_j\rangle \neq 0$  (we  abbreviate  $\langle g|\dots|g\rangle$ by $\langle\dots\rangle$; $\langle AB\rangle_{\rm s}=\langle AB+ BA\rangle/2$), and %. This anticorrelation decays over times of the order of     $\Gamma_j^{-1}$,
\beq \label{anti}
\langle \tilde{I}_j(t) I_j(t+\delta t)\rangle_{\rm s} = {\cal O}(\Gamma_j \delta t).
\eeq
    The levels $d_j$ may be implemented by quantum dots. They are assumed to be noninteracting which can be realized by small quantum dots in a  large magnetic field that can be occupied by  single electrons only. The leads coupled to the resonant levels can be emptied of and filled with electrons by applying bias voltages.

 We here introduce ${\cal E}_1$, a single observable version of the proposed entanglement measure.  For that we choose
 $\mathfrak{h}_1$ to consist of two operators:  $A_0=1$ and $A_1=Q_1=\int_\tau^\infty{dt\,\exp[-\gamma (t-\tau)]I_{1}(t)}$. We consider the regime $\Gamma_0,\Gamma_1 \ll \gamma \ll |\epsilon_0-\epsilon_1|$ and $ \tau^{-1}\approx \gamma$.  We define $C_1$ through  the states $\rho_g=|g\rangle\langle g| $ and $\rho_j= \langle g|\overline{f_j^\dag f_j^\pdag}|g\rangle^{-1} \,\overline{f_j^\pdag |g\rangle\langle g| f_j^\dag}$ ($j=0,1$). The overbar denotes a time integral, $\overline{^\pdag\dots^\pdag}= \int_0^\tau{dt\,\exp(-\gamma t)^\pdag\dots^\pdag}$.  The states $\rho_\alpha$ are  normalized such that $\lambda_\alpha(A_0) =1$ for all $\alpha$.    
 ${\cal E}_1$ of  $|g\rangle$, a measure of the entanglement of the ground state of ${\cal S}$, is now defined through Eq.\ (\ref{ent}) as the generalized entanglement of $\lambda_g$ within $C_1$ in the sense of Refs.\ \cite{Bar03,Bar04}. 
 
  The form of the current expectation values $\langle I_j\rangle \propto \langle g| f^\dag_j f^\pdag_j|g\rangle={\rm tr}\, f^\pdag_j|g\rangle\langle g|f^\dag_j$ suggests that one can project onto the states $\rho_0$ and $\rho_1$  by current measurements.  One can show that accordingly the  functionals $\lambda_{j}$ can be expressed through current correlators, $\lambda_{j}(A_1)=\langle \tilde{Q}^p_{j} Q^\pdag_1\rangle_{\rm s}/\langle \tilde{Q}^p_{j}\rangle$ with $\tilde{Q}^p_j=\int_0^\tau{dt\,\exp(-\gamma t)\tilde{I}_{j}(t)}$ \cite{unpub}.  %   ${\cal E}_1$ is thus directly observable by  measurements of electrical currents as shown 
 The $\lambda_\alpha$ that define $C_1$ expressed in terms of observables are summarized in table \ref{table1}.
       \begin{table}[htbp]
    \begin{tabular}{c|c|c|c} \hline
state $\lambda$ &density matrix & $\lambda(A_0)$ & $\lambda(A_1)$ \\ \hline
$\lambda_g $  &$ |g\rangle\langle g|$& $1$ &$\langle Q_1\rangle$  \\  \hline
$\lambda_0$ &  $\langle g|\overline{f_0^\dag f_0^\pdag}|g\rangle^{-1} \,\overline{f_0^\pdag |g\rangle\langle g| f_0^\dag}$
& $1$&   $\langle \tilde{Q}^p_0 Q_1\rangle_{\rm s}/\langle \tilde{Q}^p_0\rangle$    
   \\ \hline
       $\lambda_1 $   & $\langle g|\overline{f^\dag_1 f^\pdag_1}|g\rangle^{-1} \, \overline{f^\pdag_1 |g\rangle\langle g| f^\dag_1}$& $1$&$ \langle{\tilde {Q}}^p_1 Q_1\rangle_{\rm s}/\langle \tilde{Q}^p_1\rangle=0 $   \\ \hline
\end{tabular} 
\caption{State functionals $\lambda_j$ that define $C_1$ together with the density matrices that induce them for $\Gamma_1 \tau \to 0$.} \label{table1} 
 \end{table}
We have $\lambda_\alpha\geq0$ since all the relevant currents are positive. Also, according to Eq.\ (\ref{anti})      $\lambda_1(A_1) = 0$  in our limit $\Gamma_1 \tau \to 0$.
 One can argue \cite{unpub}, that one generically further has $\lambda_{0}(A_1)   \geq \lambda_g(A_1)$.   Hence  $\lambda_g$ is typically not extremal in $C_1$ and in that case  has the  unique representation     
 $ \lambda_g= ( \lambda_0 + \alpha \lambda_1)/(1+\alpha)$ with $ {\alpha}= \lambda_0(A_1)/\lambda_g(A_1)-1\geq 0$. With Eq.\ (\ref{ent}) we  find from this the ground state entanglement of ${\cal S}$
 \beq \label{entanglement}
 {\cal E}_1(\lambda_g) = -\frac{\alpha}{1+\alpha} \ln \frac{\alpha}{1+\alpha} -\frac{1}{1+\alpha} \ln \frac{1}{(1+\alpha )} .
 \eeq
  $\alpha$ is given by a normalized irreducible current correlator,
\beq \label{alpha}
\alpha=\frac{\langle\!\langle \tilde{Q}^p_0 Q_1\rangle\!\rangle_{\rm s}}{\langle \tilde{Q}^p_0\rangle\langle Q_1\rangle},
\eeq
where $ \langle\!\langle AB\rangle\!\rangle_{\rm s}=\langle AB\rangle_{\rm s} -  \langle A \rangle \langle B \rangle $. The correlator $\alpha$ has a well-defined limit for $v'_j\to0$, when it characterizes the ground state of ${\cal S}$ unperturbed by the measuring apparatus \cite{unpub}.  In our limits it vanishes  in the ground state of  any system of noninteracting electrons.
%  Any amount of generalized entanglement ${\cal E}_1$ that exceeds the critical value $\alpha_{\rm nonint}$   therefore unambiguously identifies many-body interactions in an  electron system.    
Thus ${\cal E}_1$  selectively characterizes ground state entanglement that is induced by many-body interactions. It is determined by cross-correlations of electrical currents,
\beq
\alpha= \int{\frac{d\omega}{2\pi} \frac{e^{-i\omega \tau}-e^{-\gamma \tau}}{1-e^{-\gamma \tau}}\frac{\gamma^2}{\omega^2+\gamma^2} \frac{\langle\!\langle \tilde{I}_0  I_1 \rangle\!\rangle_{\omega}}{\langle \tilde{I}_0\rangle\langle I_1\rangle}},
\eeq
where $\langle\!\langle \tilde{I}_0  I_1 \rangle\!\rangle_{\omega}=\int{dt\,\exp(i\omega t)\langle\!\langle \tilde{I}_0(0)  I_1(t) \rangle\!\rangle_{\rm s}}$.
There are alternative ways to measure $\alpha$ or similar correlators,  for instance using low temperature reservoirs with variable chemical potential   as  energy filters instead of resonant levels. This and a   regime of low-frequency detection $\gamma\ll \tau^{-1} \ll \Gamma_j$ will be discussed elsewhere \cite{unpub}.  
 
An intuitive interpretation of ${\cal E}_1$ in terms of     ground state  properties is possible  if the electron-electron interactions in ${\cal S}$ are absent during the current measurement. This can be achieved by  switching off the interactions at a time $\tau_s<0$ suddenly, such that ${\cal S}$ remains in its interacting ground state.   At small $\gamma$, such that on the scale $\gamma$ all ground state amplitudes depend only weakly on the single-particle energies involved, the normalized correlator $\alpha$  as defined in Eq.\ (\ref{alpha})  then takes the form  
 \beq  \label{alphan}
 \alpha=\frac{ \langle\!\langle n_0 n_1 \rangle \!\rangle}{ \langle n_0\rangle\langle  n_1 \rangle} \;\;\;\;\;  {\rm for } \;\; f_j=d_j .
 \eeq 
For a finite-sized system in the  limit  $\tau_s\Gamma_j \to -\infty$   $n_j=\psi^\dag_{k_j} \psi^\pdag_{k_j} $    is the occupation number of noninteracting eigenmodes $\psi_{k_j}$ of ${\cal S} $ with quantum numbers $k_j$ and noninteracting energies $\epsilon_{k_j}= \epsilon_j$ \cite{footnote2}.   $  {\cal E}_1$ in that case thus quantifies  correlations between fermion occupation numbers in the ground state of ${\cal S}$, similarly to the     entanglement measures  proposed in  Refs.\ \cite{Zan02,Shi03,Shi04}. %These correlations are typically induces by particle-hole pairs in the interacting ground state.
   %These  correlations   are typically induced by interactions through the generation of electron-hole pairs: while the hole within such a pair reduces the electron occupation number $n_0$ at one energy $\epsilon_0$, the corresponding electron reduces the hole occupation number $1-n_1$ at another energy $\epsilon_1$.
%This mechanism evidently produces positive correlations between $n_0$ and $1-n_1$, such that $\alpha>0$.  We conclude that $ \lambda_0(A_1)= \lambda_g(A_1)(1+\alpha)$  is indeed generically larger than $\lambda_g(A_1)$ for an interacting system, as claimed above. 
Such correlations are induced by interactions typically through the creation of particle-hole pairs.
   Extensions of  $  {\cal E}_1$ that capture ground state entanglement due to multiple particle-hole pairs  will be discussed in \cite{unpub}.
A sudden switching off of interactions is possible in cold gases of fermionic atoms \cite{Reg04}.
% Also certain electronic structures may be regarded as effectively noninteracting during the correlation measurement, as discussed below. 
Electronic systems, however, may be regarded as noninteracting during the current measurement only in special situations, as discussed below.   %$\alpha$ can then be expressed in terms of overlaps of $\psi|g\rangle$ or $\psi^\dag |g\rangle$ with excited states of ${\cal S}$, preferably with energies differing by $\epsilon_0$ or $\epsilon_1$ from the ground state energy.  
In general they have no accessible noninteracting counterpart. There is then no natural set of single-particle modes and
 ${\cal E}_1$ lacks a comparable intuitive interpretation.  It is a generalization of occupation number entanglement to such cases. 
 %It probes interaction-induced differences in the amplitudes for two-particle excitations versus single particle excitations at selectable energies.

\begin{figure} 
\includegraphics[width=4.5cm]{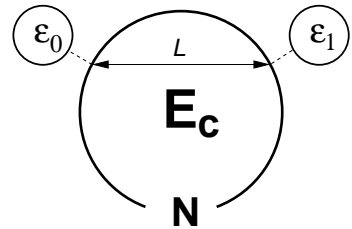}
%\vspace{4cm}
\caption{  Large quantum dot opened up by $N$ conduction channels coupled to two resonant levels implemented by small dots. }  \label{fig2}
\end{figure}

 We now evaluate ${\cal E}_1$  for two typical electronic nanostructures. 
 Since we are interested in their ground state entanglement we take the limit of zero temperature  in all examples. First  we consider an open quantum dot with charging interaction. 
 We assume that the charging energy $E_c$ of the dot is large and address the regime $ \Gamma_j \ll  \gamma \ll |\epsilon_j| \ll E_c$, $ \tau^{-1}\approx \gamma$, and $v'_j\ll v_j,\tilde{v}_j$.             
We additionally require a small level spacing $\Delta\ll \Gamma_j$ and an in-plane magnetic field that renders the electrons effectively spinless.
% A continuum description of the states in the quantum dot is then appropriate.
We probe the quantum dot at two points at a distance $L$ from each other by two resonant levels $0$ and $1$  contacted with empty and occupied reservoirs respectively, as shown in Fig.\ \ref{fig2}. We choose  $\epsilon_0<0$ and $\epsilon_1>0$ relative to the  Fermi energy $\mu_{\rm QD}=0$ of the dot.   The quantum dot is opened by point contacts with a large total number of channels $N\gg 1$.  The effects of the  charging interaction can then be treated perturbatively within the incoherent model of quantum dots \cite{Ale02}.  We find at $L=0$ to leading order in our limits
\beq \label{EQD}
  {\cal E}^{\rm QD}_1 =  \frac{1}{N} \frac{\bar{\Gamma}}{\epsilon_1-\epsilon_0} \left(\ln N \frac{\epsilon_1-\epsilon_0} {\bar{\Gamma}}+1\right)       ,
\eeq 
where $\bar{\Gamma}^{-1} = (\Gamma_0^{-1}+\Gamma_1^{-1})/2$.  In this structure  not only the leading order statistical correlations $\alpha^{\rm QD}_{\rm nonint}=-\Gamma_0\Gamma_1/(\epsilon_1-\epsilon_0)^2$, but  also the interaction-induced correlations $\alpha^{\rm QD}=\bar{\Gamma}/N(\epsilon_1-\epsilon_0)$  become small for $\bar{\Gamma} \ll \epsilon_1-\epsilon_0$. The latter ones, however,  dominate    and thus unambiguously identify many-body interactions for $(\Gamma_0+\Gamma_1)/2(\epsilon_1-\epsilon_0)\ll1/N$. This is the regime of validity of Eq.\ (\ref{EQD}). It lies well within the range of typical experimental parameters.  $\alpha^{\rm QD}$ decays over distances $L\simeq {\rm min} \{v_F/\Gamma_j, l_{\rm in}\}$, where $v_F$ is the Fermi velocity and $l_{\rm in}$  the  inelastic length, and it acquires an oscillatory contribution at $L \simeq v_F/|\epsilon_1-\epsilon_0|$ \cite{unpub}. It persists up to temperatures of the order of $|\epsilon_j|$.  %Additionally these interaction-induced correlations can be experimentally distinguished from the noninteracting contribution by their dependence on $N$. 
 Eq.\ (\ref{alphan}) cannot be used to interpret ${\cal E}_1^{\rm QD}$   since the charging interaction cannot be switched off.  Nevertheless,  $\alpha^{\rm QD} \propto (\epsilon_1-\epsilon_0)^{-1}$ reflects the typical energy-dependence of the probability of particle-hole excitations in effectively one-dimensional interacting electron systems.  This illustrates that in particular the dependence of ${\cal E}_1$ on $\epsilon_0$ and $\epsilon_1$ contains valuable information about the ground state  of an interacting fermion system. Ref.\ \cite{Cra04} reports an experiment on a setup very similar to the one shown  in Fig.\ \ref{fig2}.  The   measurement of the proposed  many-fermion entanglement in this structure should thus be experimentally feasible.

\begin{figure} 
\includegraphics[width=5.5cm]{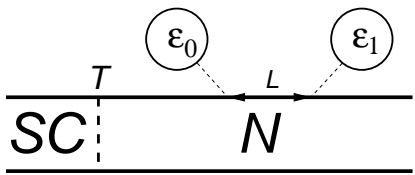}
%\vspace{4cm}
\caption{  Normal metal (N) tunnel-coupled (transmission $T\ll 1$) to a superconductor (SC). }  \label{fig3}
\end{figure}

As a second example we consider  a structure where interactions may effectively be switched off during the current measurement: a piece of  normal metal with well-screened interactions weakly coupled to a BCS superconductor, as shown in Fig.\ \ref{fig3}.  The superconductor induces pair-correlations in the normal metal through the so-called proximity effect that result in a nonzero many-fermion entanglement ${\cal E}_1$. If the measurement is performed in the noninteracting part of the structure at a  sufficiently large distance from the superconductor, where the pairing interaction is present, one expects that  ${\cal E}_1$ can be interpreted  with the help of Eq.\ (\ref{alphan}). We will confirm this intuition in Ref.\ \cite{unpub}:  for certain (not necessarily lowest energy) states one can show that the normalized correlator $\alpha^{\rm SC}$ is indeed in precise correspondence with  occupation number correlations in the spirit of Eq.\ (\ref{alphan}).    Also in this example much of the information content of $\alpha^{\rm SC}$ is in its dependence on the level energies $\epsilon_j$:    $\alpha^{\rm SC}$ is maximal for $\epsilon_0=-\epsilon_1$ (the Fermi energy of the superconductor is chosen $\mu_{\rm SC}=0$) and it decays  for  $|\epsilon_0+\epsilon_1|>\Gamma_j$ \cite{unpub}.

Other entanglement measures   based on current correlations   in nanostructures have been put forward in Refs.\ \cite{Bur00}. These entanglement measures are designed to quantify two-particle correlations. They do not distinguish interaction-induced correlations from statistical ones. The entanglement captured by these measures is  present also in certain noninteracting fermion systems \cite{Bee03}. This contrasts with the many-fermion entanglement proposed here which vanishes in noninteracting ground states. It singles out correlations due to many-body interactions.

In conclusion, we have proposed a measure of many-fermion  entanglement  based on a generalized notion of entanglement developed in Refs.\ \cite{Bar03,Bar04}. It quantifies ground state correlations that are induced by many-body interactions  and it is experimentally accessible through cross-correlation measurements of currents through resonant fermion levels. % It quantifies correlations of single particle state occupation numbers with selectable energies in the spirit of Refs.\ \cite{Zan02,Shi03} if one has control over the many-body interaction strength. 
It measures a generalization of occupation number entanglement  \cite{Zan02,Shi03}. 
%: it vanishes for noninteracting systems and probes interaction-induced differences in the amplitudes for two-particle excitations versus single particle excitations at selectable energies. 
With two examples we have illustrated the introduced entanglement measure and highlighted aspects of interaction-induced ground state correlations  that it serves to characterize. The experimental techniques that its measurement requires are all existing in electronic nanostructures. It thus promises to be a systematic way of studying interacting fermion systems, both, theoretically as well as experimentally.

 The author thanks P.\ W.\ Brouwer and D.\ M.\ Eigler for valuable discussions and acknowledges support   by the Packard Foundation.


\begin{thebibliography}{99}
\bibitem{footnote1}
 Since we are only interested in the ground state of ${\cal S}$ we employ here a  measure of pure state entanglement  (pure in an absolute sense, with respect to all operators and states of ${\cal S}$).
\bibitem{footnote2}
Without the assumptions made in the main text Eq.\ (\ref{alphan}) is in our limit  $\Gamma_j \ll \gamma \ll |\epsilon_0-\epsilon_1|$, $\gamma \approx \tau^{-1}$ generically valid with an  $n_j\propto \sum_{kk'} \psi^\dag_{k'}\psi^{\phantom{\dagger}}_{k}/(\epsilon_j-\epsilon_{k}-i \Gamma_j/2)(\epsilon_j-\epsilon_{k'}+i \Gamma_j/2)(\epsilon_k-\epsilon_{k'}-i \gamma)$, provided that the overlap of the mode $\psi_j$ (coupled to $d_j$) with the noninteracting eigenstates $\psi^\pdag_{k}$  varies weakly on the scale $\gamma$ \cite{unpub}.
\end{thebibliography}
\end{document}